# Energy levels of $In_xGa_{1-x}As$/GaAs quantum dot laser with different sizes

Esfandiar Rajaei[*] and Mahdi Ahmadi Borji[**]

Department of Physics, Faculty of Science, University of Guilan, P.O. Box No: 41335-19141, Rasht, Iran

*Corresponding author, E-mail: Raf404@guilan.ac.ir, **Email: Mehdi.p83@gmail.com

**Abstract**

*In this paper, we have studied the strain, bandedge, and energy levels of cubic InGaAs quantum dots (QDs) surrounded by GaAs. It is shown that overall strain value is larger in InGaAs-GaAs interfaces, and also in smaller QDs. Also, it is proved that conduction and valence band-edges and electron-hole levels are size dependent; larger QD sizes appeared to result in the lower recombination energies. Moreover, more number of energy levels separates from the continuum states of bulk GaAs and comes down into the QD separate levels. In addition, degeneracy of eigenvalues was found to be subjected to change by size variation. Our results coincide with former similar researches.*

*Keywords: strain, bandedge, engineering energy levels, quantum dot size, QD laser*

## I.    INTRODUCTION

Semiconductor lasers have found many applications, and among many types of them, Quantum dot lasers have found a special place in new life due to their interesting characteristics arising from their discrete energy levels. Effects of various factors such as QD size [1, 2], percentage of constituent elements of the QD [3], substrate index [4], strain [5, 6], usage temperature [7-10], wetting layer (WL), and distribution of QDs are shown to be important in the energy levels and performance of quantum dot lasers. Thus, finding the effect of these factors can be instructive in optimizing the laser performance. QD size effects are interesting and important, since it can change recombination energies and carrier relaxation and recombination times [11].

Quantum dots have been the focus of many researches due to their optical properties arising from the dimensional confinement of carriers [12-15]. They have found many applications in semiconductor lasers and optical amplifiers [16-19]. $In_xGa_{1-x}As$/GaAs devices are now widely used in laser devices [15, 20-24]. So, a ubiquitous view of the energy states, bandedges, strain, and other features, and their variation by QD size can be helpful.

In semiconductor hetero-structures which contain more than one material, energy states appear to be more complex than bulk samples due to the significant role of strain. Strain tensor depends on lattice mismatch, elastic properties of neighbor materials, and geometry of the QD [25]. This research represents a quantum numerical study of the energy states, band structure, and strain tensor of $In_{0.2}Ga_{0.8}As$ QDs grown on GaAs substrate.

The rest of this paper is organized as follows: section II explains the numerical model; results and discussions are presented in section III; finally, we make a conclusion in section IV.

## II. NUMERICAL MODEL

In self-assembled QD growth, a WL with a few molecular layers is grown and millions of QDs are formed, each with a random shape and size. QDs are finally covered by a cap layer. Many shapes can be approximated for QDs, namely, cylindrical, cubic, lens shape, pyramidal [26], etc. For simplicity, the QDs are assumed here to be cubic and far enough to avoid any effects by neighbor QDs. The one-band effective mass approach is used in solving the Schrödinger equation, and the Poisson's equation was solved numerically in a self-consistent manner.

Figure 1 shows the cross-section of a $4 \times 4 \times 4 nm^3$ cubic $In_{0.2}Ga_{0.8}As$ QD surrounded by GaAs. The substrate and cap thickness are assumed here to be 20nm, and the wetting layer to be $0.5 nm$. This structure is grown on (001) substrate index. The growth-direction is along z-axis. The unstructured mesh is used for the system in which smaller meshes are included inside the QD region.

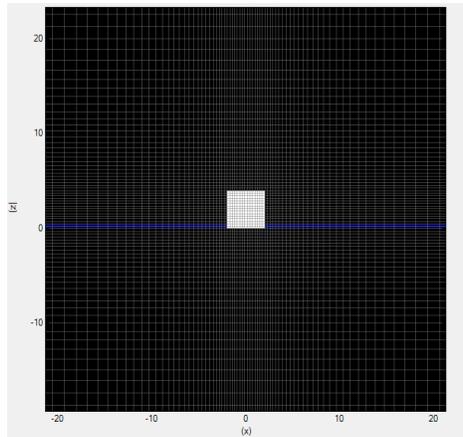

Fig. 1: Profile of a cubic InGaAs QD of $4nm \times 4nm \times 4nm$ on $20nm$ thick GaAs substrate and $0.5nm$ wetting layer. The meshes are seen in this figure.

When changing QD size, all the sides change simultaneously and the cubic shape is fixed. Also, the cell volume changes since cap and substrate index are fixed.

## III. RESULTS AND DISCUSSION

Strain is defined as the summation of all infinitesimal length increases relative to the instantaneous lengths ($\varepsilon_L = \sum \Delta L_t / L_t$). Thus, by taking into account length changes in all dimensions, one achieves a strain tensor

$$\varepsilon_{ij} = \frac{1}{2}\left(\frac{du_i}{dr_j} + \frac{du_j}{dr_i}\right) \qquad , \quad i,j \equiv x,y,z \tag{1}$$

where $du_i$ is the length variation in i-th direction, and $r_j$ is the length in direction $j$ [4].

Fig. 2 illustrates the two non-zero elements of the strain tensor, namely, $\varepsilon_{xx}$ and $\varepsilon_{zz}$ for two different QD sizes. As it is viewed, in both directions, strain tensor is subjected to change in interfaces. However, the near points show to have different strains as well.

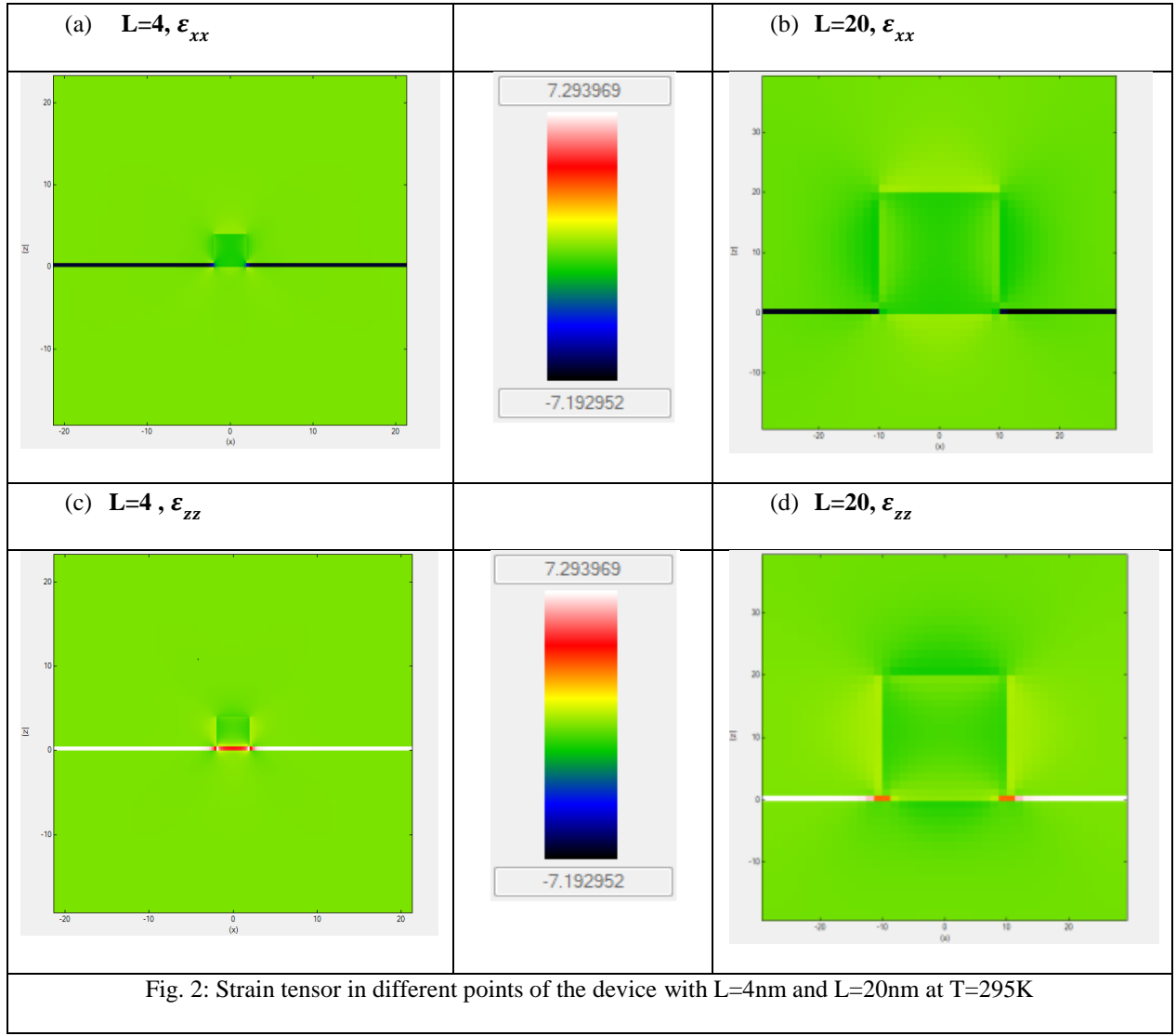

Fig. 2: Strain tensor in different points of the device with L=4nm and L=20nm at T=295K

In Fig. 3 all elements of strain tensor are plotted along z-direction and at the middle cross-section of the structure. Here $\varepsilon_{11}$ and $\varepsilon_{22}$ coincide on one curve, and $\varepsilon_{13}$, $\varepsilon_{23}$, and $\varepsilon_{12}$ coincide on another one with zero value, showing almost no effect of mismatches in one direction on strain exerting on atoms in another direction. Moreover, it can be argued that the existence of indium in one side of interfaces leads to a jump in the strain tensor for $\varepsilon_{11}$, $\varepsilon_{22}$ and $\varepsilon_{33}$ meaning a stretch in GaAs and squeeze in InGaAs lattice constant.

In z-direction, as it is observed, strain experienced in interfaces is different for different QD sizes; the value is more for smaller QDs which informs of more stretch exerting on few number of atoms existing in a smaller QD. To explain, we notice that lattice constant of GaAs and InAs is 0.565325nm and 0.60577nm respectively, and it increases almost linearly by indium percentage [26]. Strain is discussed in [27] that can be due to 7% mismatch of lattice constants of GaAs and InAs [28].

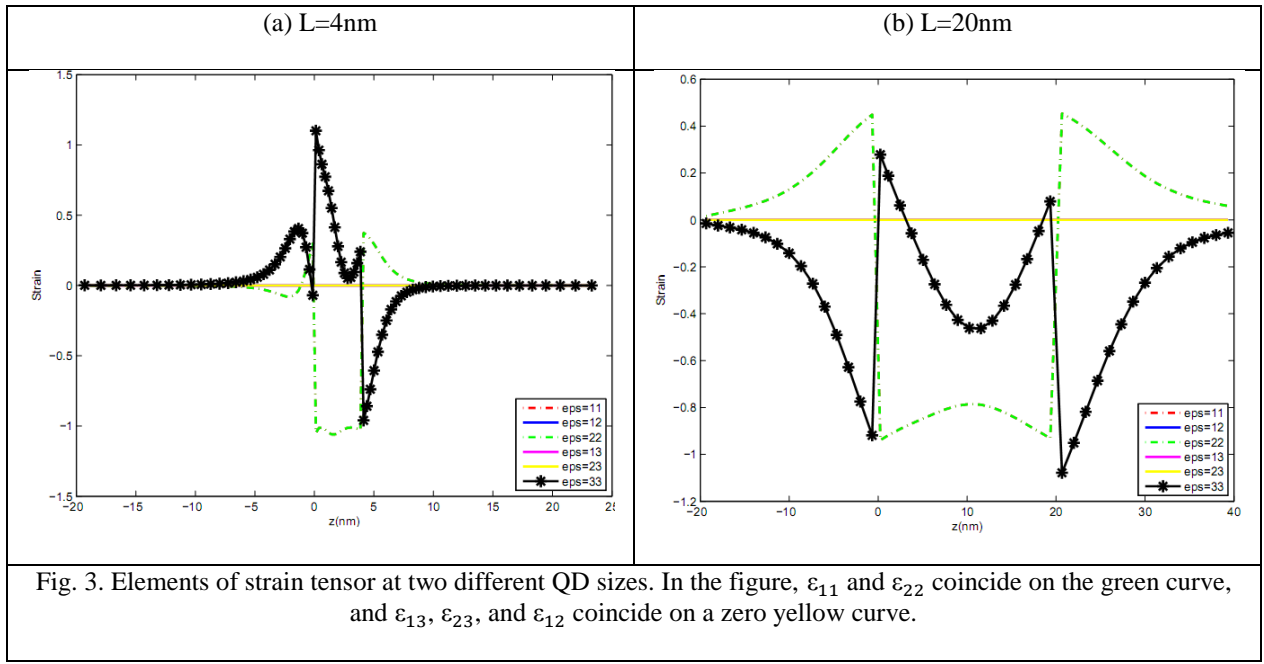

Fig. 3. Elements of strain tensor at two different QD sizes. In the figure, $\varepsilon_{11}$ and $\varepsilon_{22}$ coincide on the green curve, and $\varepsilon_{13}$, $\varepsilon_{23}$, and $\varepsilon_{12}$ coincide on a zero yellow curve.

Fig. 3 shows the Γ and Heavy-Hole (HH) band-edges of each point of QDs with different sizes in x-z plane in the middle cross-section of the structure. As it is observed, both conduction and valence band-edges have been fully subjected to change by size effect. Also, the cap layer bandedges have been subjected to change in the points close to the QD. Comparison of Fig 3(a) with 3(c) or Fig. 3(b) with 3(d) shows that both electron- and hole-bandedges of the QD are sensitive to size.

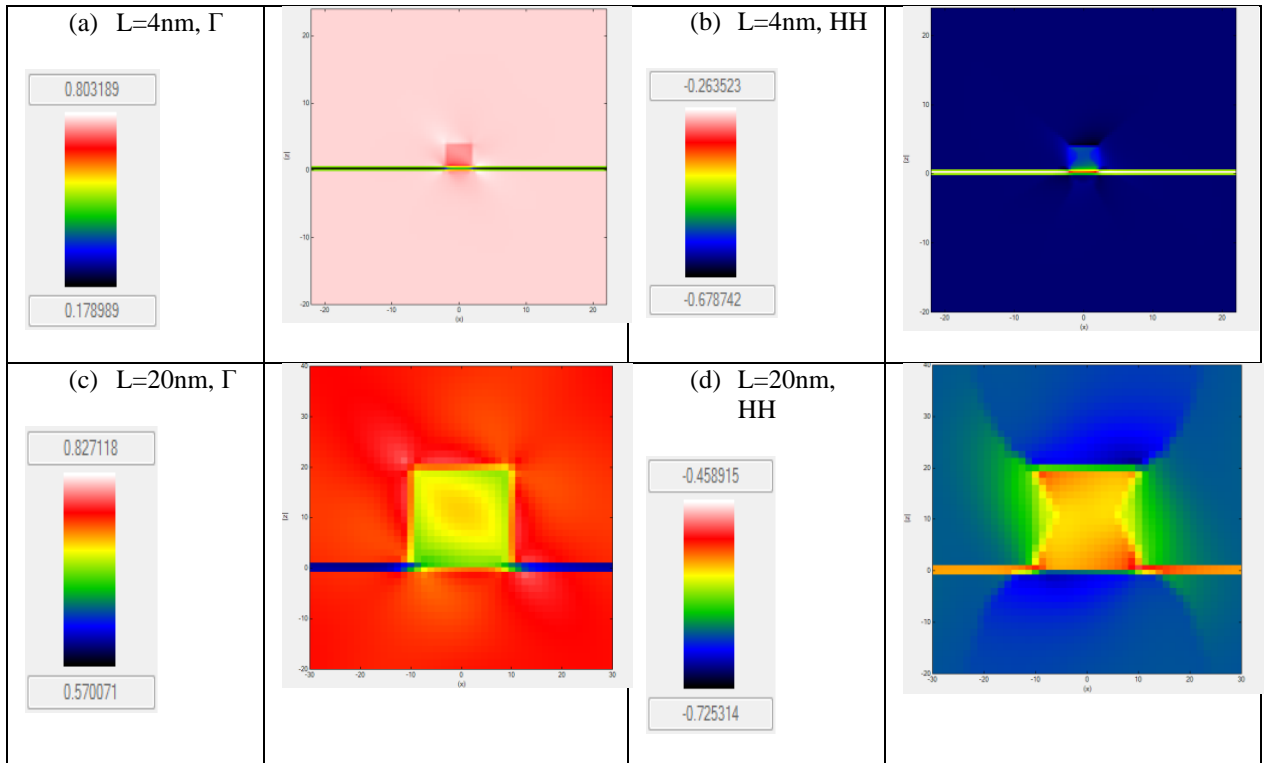

Fig. 3. Valence and conduction band-edges in x-z plane in the cross-section of the QD. Note the numbers related to colors beside each figure.

In Fig. 4 conduction and valence band-edges in z-direction, and three first allowed energy states of electrons and holes are shown. As it is clear, QDs of side 4nm has no allowed energy state for lectrons into the QD. But enlargement of QD side to 14nm leads to a lowered electronic state which lay into the QD. More increase of QD size results in the more separated energy levels laid into the QD. Moreover, the recombination energies have decreased by size. Other states are among continuous states of the GaAs. Since in a laser device, the photons are emitted from the separate energies of the QD, the laser wavelength is expected to elongate in larger QDs. Some similar results can be seen in [1, 2, 27] in which the size dependence is confirmed. In addition, energy gap for bulk InAs and GaAs are 0.36eV and 1.43eV respectively [29], but clearly, it changes here by size restriction.

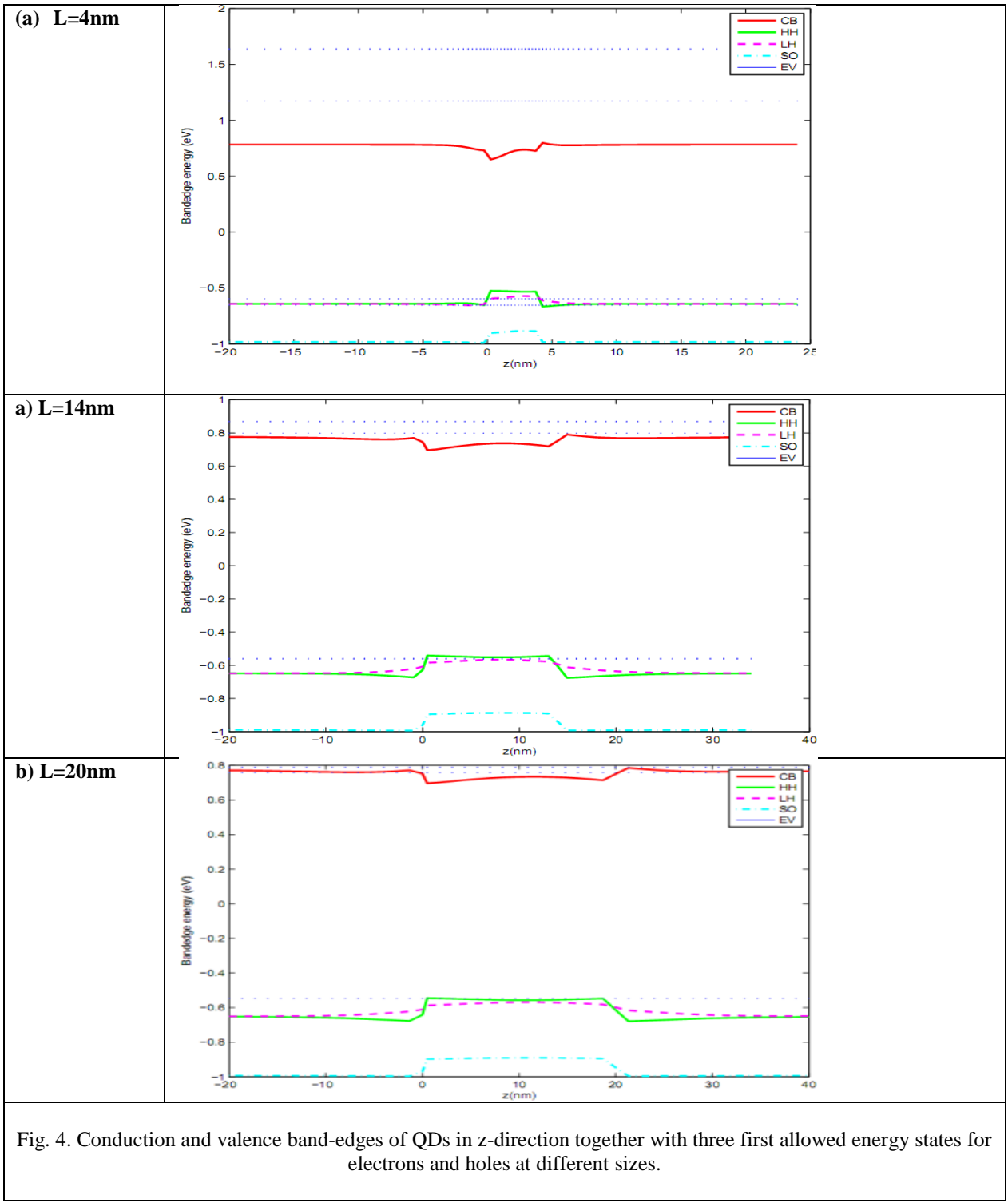

Fig. 4. Conduction and valence band-edges of QDs in z-direction together with three first allowed energy states for electrons and holes at different sizes.

In addition, in Table 1, the first five eigenvalues of electron- and hole-states of different QD sizes are written. Clearly, size change has changed the degeneracy of both electron and hole states as well as the recombination energies.

| QD Size (nm) | Level No | Electron energy (eV) | Hole energy (eV) | Recombination energy | Status |
|---|---|---|---|---|---|
| 4 | 1 | 1.17 | -0.59 | 1.76 | **Among continuum states** |
| 4 | 2 | 1.63 | -0.65 | 2.28 | **Among continuum states** |
| 4 | 3 | 1.64 | -0.65 | 2.30 | **Among continuum states** |
| 4 | 4 | 1.64 | -0.66 | 2.30 | **Among continuum states** |
| 4 | 5 | 2.10 | -0.71 | 2.81 | **Among continuum states** |
| 14 | 1 | 0.79 | -0.55 | 1.35 | **Into QD separate states** |
| 14 | 2 | 0.86 | -0.56 | 1.43 | **Among continuum states** |
| 14 | 3 | 0.86 | -0.56 | 1.43 | **Among continuum states** |
| 14 | 4 | 0.87 | -0.57 | 1.44 | **Among continuum states** |
| 14 | 5 | 0.94 | -0.57 | 1.51 | **Among continuum states** |
| 20 | 1 | 0.75 | -0.54 | 1.30 | **Into QD separate states** |
| 20 | 2 | 0.78 | -0.54 | 1.33 | **Into QD separate states** |
| 20 | 3 | 0.78 | -0.55 | 1.34 | **Into QD separate states** |
| 20 | 4 | 0.79 | -0.55 | 1.35 | **Into QD separate states** |
| 20 | 5 | 0.82 | -0.55 | 1.38 | **Into QD separate states** |

**Table 1:** First five eigenvalues of electron and hole for different QD sizes

## IV. CONCLUSION

We studied the band structure and strain tensor of $In_{0.2}Ga_{0.8}As$ quantum dots grown on GaAs substrate by numerical solutions. It was shown that the total strain value is larger in the interfaces, and also in smaller QD sizes. The conduction and valence band-edges and electron-hole levels were found to be dependent on QD size as well; larger sizes resulted in the lower (higher) energy of electrons (holes). Thus, the recombination energies decreased in larger QDs. In addition, more number of energy levels separated from the continuum states of bulk GaAs and came down into the QD separate levels. Moreover, it was observed that degeneracy of levels was subjected to change by size variation. These results had a good consonance with Pryor et al results [2, 5].

**Acknowledgement**

The authors give the sincere appreciation to Dr. S. Birner for providing the advanced 3D Nextnano++ simulation program [31] and his instructive guides. We would like to thank numerous colleagues, namely, Prof. S. Farjami Shayesteh and S. M. Rozati, Dr. Hamed Behzad, K. Kayhani, and Y. Yekta Kia for sharing their points of view on the manuscript.

**References:**

1. Baskoutas, S. and A.F. Terzis, *Size-dependent band gap of colloidal quantum dots.* Journal of Applied Physics, 2006. **99**(1): p. 013708.
2. Pryor, C., *Eight-band calculations of strained InAs/GaAs quantum dots compared with one-, four-, and six-band approximations.* Physical Review B, 1998. **57**(12): p. 7190-7195.
3. Shi, Z., et al., *Influence of V/III ratio on QD size distribution.* Frontiers of Optoelectronics in China, 2011. **4**(4): p. 364-368.
4. Povolotskyi, M., et al., *Tuning the piezoelectric fields in quantum dots: microscopic description of dots grown on (N11) surfaces.* Nanotechnology, IEEE Transactions on, 2004. **3**(1): p. 124-128.
5. Pryor, C.E. and M.E. Pistol, *Band-edge diagrams for strained III\char21{}V semiconductor quantum wells, wires, and dots.* Physical Review B, 2005. **72**(20): p. 205311.
6. Shahraki, M. and E. Esmaili, *Computer simulation of quantum dot formation during heteroepitaxial growth of thin films.* Journal of Theoretical and Applied Physics, 2012. **6**(1): p. 1-5.
7. CHEN, S.-H. and J.-L. XIAO, *TEMPERATURE EFFECT ON IMPURITY-BOUND POLARONIC ENERGY LEVELS IN A PARABOLIC QUANTUM DOT IN MAGNETIC FIELDS.* International Journal of Modern Physics B, 2007. **21**(32): p. 5331-5337.
8. Kumar, D., C.M.S. Negi, and J. Kumar, *Temperature Effect on Optical Gain of CdSe/ZnSe Quantum Dots*, in *Advances in Optical Science and Engineering*, V. Lakshminarayanan and I. Bhattacharya, Editors. 2015, Springer India. p. 563-569.
9. Narayanan, M. and A.J. Peter, *Pressure and Temperature Induced Non-Linear Optical Properties in a Narrow Band Gap Quantum Dot.* Quantum Matter, 2012. **1**(1): p. 53-58.
10. Rossetti, M., et al., *Modeling the temperature characteristics of InAs/GaAs quantum dot lasers.* Journal of Applied Physics, 2009. **106**(2): p. 023105.
11. Heitz, R., et al., *Energy relaxation by multiphonon processes in InAs/GaAs quantum dots.* Physical Review B, 1997. **56**(16): p. 10435-10445.
12. Markéta ZÍKOVÁ, A.H., *Simulation of Quantum States in InAs/GaAs Quantum Dots.* NANOCON 2012. **23**(25): p. 10.
13. Ma, Y.J., et al., *Factors influencing epitaxial growth of three-dimensional Ge quantum dot crystals on pit-patterned Si substrate.* Nanotechnology, 2013. **24**(1): p. 015304.
14. DANESH KAFTROUDI, Z. and E. RAJAEI, *SIMULATION AND OPTIMIZATION OF OPTICAL PERFORMANCE OF INP-BASED LONGWAVELENGTH VERTICAL CAVITY SURFACE EMITTING LASER WITH SELECTIVELY TUNNEL JUNCTION APERTURE.* JOURNAL OF THEORETICAL AND APPLIED PHYSICS (IRANIAN PHYSICAL JOURNAL), 2010. **4**(2): p. 12-20.
15. Nedzinskas, R., et al., *Polarized photoreflectance and photoluminescence spectroscopy of InGaAs/GaAs quantum rods grown with As(2) and As(4) sources.* Nanoscale Research Letters, 2012. **7**(1): p. 609-609.
16. Bimberg, D., et al., *Quantum dot lasers: breakthrough in optoelectronics.* Thin Solid Films, 2000. **367**(1–2): p. 235-249.
17. Gioannini, M., *Analysis of the Optical Gain Characteristics of Semiconductor Quantum-Dash Materials Including the Band Structure Modifications Due to the Wetting Layer.* IEEE Journal of Quantum Electronics, 2006. **42**(3): p. 331-340.
18. KAFTROUDI, D., et al., *Thermal simulation of InP-based 1.3 μm vertical cavity surface emitting laser with AsSb-based DBRs*. Vol. 284. 2011, Amsterdam, PAYS-BAS: Elsevier. 11.
19. Asryan, L.V. and S. Luryi, *Tunneling-injection quantum-dot laser: ultrahigh temperature stability.* Quantum Electronics, IEEE Journal of, 2001. **37**(7): p. 905-910.

20. Woolley, J.C., M.B. Thomas, and A.G. Thompson, *Optical energy gap variation in GaxIn1–x As alloys.* Canadian Journal of Physics, 1968. **46**(2): p. 157-159.
21. Hazdra, P., et al., *Optical characterisation of MOVPE grown vertically correlated InAs/GaAs quantum dots.* Microelectronics Journal, 2008. **39**(8): p. 1070-1074.
22. Fali, A., E. Rajaei, and Z. Kaftroudi, *Effects of the carrier relaxation lifetime and inhomogeneous broadening on the modulation response of InGaAs/GaAs self-assembled quantum-dot lasers.* Journal of the Korean Physical Society, 2014. **64**(1): p. 16-22.
23. Yekta Kiya, Y., E. Rajaei, and A. Fali, *Study of response function of excited and ground state lasing in InGaAs/GaAs quantum dot laser.* J. Theor. Phys. , 2012. **1**: p. 246-256.
24. Azam Shafieenezhad, E.R., , Saeed Yazdani, *The Effect of Inhomogeneous Broadening on Characteristics of Three-State Lasing Ingaas/Gaas Quantum Dot Lasers.* International Journal of Scientific Engineering and Technology, 2014. **3**(3): p. 297- 301.
25. Trellakis, A., et al., *The 3D nanometer device project nextnano: Concepts, methods, results.* Journal of Computational Electronics, 2006. **5**(4): p. 285-289.
26. Qiu, D. and M.X. Zhang, *The preferred facet orientation of GaAs pyramids for high-quality InAs and InxGa1−xAs quantum dot growth.* Scripta Materialia, 2011. **64**(7): p. 681-684.
27. Jiang, H. and J. Singh, *Conduction band spectra in self-assembled InAs/GaAs dots: A comparison of effective mass and an eight-band approach.* Applied Physics Letters, 1997. **71**(22): p. 3239-3241.
28. Zhao, C., et al., *Quantum-dot growth simulation on periodic stress of substrate.* The Journal of Chemical Physics, 2005. **123**(9): p. -.
29. Bratkovski, A. and T.I. Kamins, *Nanowire-Based Light-Emitting Diodes and Light-Detection Devices With Nanocrystalline Outer Surface.* 2010, Google Patents.
30. Xu, S., et al., *Effects of size polydispersity on electron mobility in a two-dimensional quantum-dot superlattice.* Physical Review B, 2014. **90**(14): p. 144202.
31. Birner, S., et al., *nextnano: General Purpose 3-D Simulations.* Electron Devices, IEEE Transactions on, 2007. **54**(9): p. 2137-2142.